\documentclass[fleqn,usenatbib]{mnras}

\usepackage{newtxtext,newtxmath}

\usepackage[T1]{fontenc}

\DeclareRobustCommand{\VAN}[3]{#2}
\let\VANthebibliography\thebibliography
\def\thebibliography{\DeclareRobustCommand{\VAN}[3]{##3}\VANthebibliography}

\usepackage{graphicx}	
\usepackage{amsmath}	
\usepackage[utf8]{inputenc}
\usepackage[english]{babel}

\usepackage[dvipsnames]{xcolor}
\definecolor{mypink1}{RGB}{255, 127, 127}
\definecolor{mypurple}{RGB}{179, 0, 255}

\title[ ]{Understanding the "Feeble Giant" Crater II with tidally stretched Wave Dark Matter}

\author[A. Pozo et al.]{Alvaro Pozo$^{1,2}$, Tom Broadhurst$^{1,2,3}$,
Razieh Emami$^{4}$,
George Smoot$^{2,5,6,7,8}$\\
$^{1}$Department of Physics, University of the Basque Country UPV/EHU, E-48080 Bilbao, Spain;\\ email:alvaro.pozolarrocha@bizkaia.eu; tom.j.broadhurst@gmail.com;\\
$^{2}$DIPC, Basque Country UPV/EHU, E-48080 San Sebastian, Spain\\
$^{3}$ Ikerbasque, Basque Foundation for Science, E-48011 Bilbao, Spain\\
$^{4}$ Center for Astrophysics $\vert$ Harvard \& Smithsonian, 60 Garden Street, Cambridge, MA 02138, USA \\
$^{5}$\textit{Institute for Advanced Study and Department of Physics, IAS TT \& WF Chao Foundation Professor, Hong Kong University of Science and Technology, Hong Kong}\\
$^{6}$\textit{Energetic Cosmos Laboratory, Nazarbayev University, Nursultan, Kazakhstan}\\
$^{7}$\textit{Physics Department,
University of California at  Berkeley CA 94720 Emeritus}\\
$^{8}$\textit{Paris Centre for Cosmological Physics, APC, CNRS/IN2P3, CEA/lrfu,Universit\'{e} de Paris, 10, rue Alice Domon et Leonie Duquet,} \\
\textit{75205 Paris CEDEX 13, France  Emeritu}
}
\date{Accepted XXX. Received YYY; in original form ZZZ}

\pubyear{2021}

\begin{document}

\label{firstpage}
\pagerange{\pageref{firstpage}--\pageref{lastpage}}
\maketitle

\begin{abstract}

The unusually large "dwarf" galaxy Crater II, with its small velocity dispersion, $\simeq 3$ km/s, defies expectations that low mass galaxies should be small and dense. We combine the latest stellar and velocity dispersion profiles finding Crater II has a prominent dark core of radius $\simeq 0.71^{+0.09}_{-0.08}$ kpc, surrounded by a low density halo, with a transition visible between the core and the halo. We show that this profile matches the distinctive core-halo profile predicted by "Wave Dark Matter" as a Bose-Einstein condensate, $\psi$DM, where the ground state soliton core is surrounded by a tenuous halo of interfering waves, with a marked density transition predicted between the core and halo. Similar core-halo structure is seen in most dwarf spheroidal galaxies (dSph), but with smaller cores, $\simeq 0.25$ kpc and higher velocity dispersions, $\simeq 9$km/s, and we argue here that Crater II may have been a typical dSph that has lost most of its halo mass to tidal stripping, so its velocity dispersion is lower by a factor of 3 and the soliton is wider by a factor of 3, following the inverse scaling required by the Uncertainty Principle. This tidal solution for Crater II in the context of $\psi$DM, is supported by its small pericenter of $\simeq 20$ kpc established by Gaia, implying significant tidal stripping of Crater II by the Milky Way is expected.
 
\end{abstract}






\section{Introduction}



A surprising diversity of dwarf galaxies has been steadily uncovered over the past decade, with many ``ultra faint dwarfs" \citep{Moskowitz2020,Pakdil2018,Munoz2018,Koposov2015,Read2006} that are much smaller and denser than the well studied class of dwarf spheroidal galaxies (dSph), and other ``ghostly" dwarfs of very low surface brightness that are large and lower in density. The aptly named ``feeble giant", Crater II, has been particularly puzzling, with its dwarf-like velocity dispersion of only $\simeq 3$ km/s \citep{Torrealba2016,Caldwell2017} and large size, over $\simeq 2$ kpc, that strains the dwarf definition in terms of size. Despite the low velocity dispersion, the mass-to-light ratio of Crater II is large, $M/L\simeq 30$ \citep{Caldwell2017,Ji2021,Sanders2018} and appears to have a shallow, cored profile \citep{Ji2021,Sanders2018} and an old, but not ancient stellar population, dated to $\simeq 10$ Gyrs \citep{Torrealba2016,Caldwell2017}, resembling in these respects the common class of dwarf Spheroidal galaxy (dSph) for which dark matter-dominated cores are commonly claimed. However, both the stellar surface brightness and the velocity dispersion of Crater II are much lower than the typical dSph dwarfs.

Recently it has become clear that the dark cores of the dSph galaxies match well an essential prediction for dark matter as a Bose-Einstein condensate, for which a standing wave, soliton forms a prominent core in every galaxy \citep{Schive2014}. The first cosmological simulations in this context have revealed pervasive wave structure on the de Broglie scale, termed $\psi$DM, \citep{Schive2014, Schive2014b, Schwabe2016,Hui2017,Mocz2017}. This includes a stable soliton core formed at the center of each galaxy, corresponding to the ground state \citep{Schive2014}, surrounded by a halo of excited states. A boson mass of $\simeq 10^{-22}$ eV provides cores of $\simeq 0.5$ kpc typical of the dwarf spheroidal \citep{Pozo2020,idm2018,Schive2014b}, where the width of the soliton core is set simply by the de Broglie wavelength and is seen to vary between galaxies in the simulations, being larger for lower mass galaxies, reflecting their lower momentum. Here, the local de Broglie scale is $\lambda_{dB} = h/(m_\psi \sigma)$, where $\sigma$ is the local velocity dispersion that sets the momentum scale together with the boson mass $m_\psi$.

This is a scalar field interpretation of DM described simply by a coupled Schrodinger-Poisson equation for the mean-field behavior that evolves only under self-gravity, for which the boson mass is the only free parameter. The most unique prediction is that a prominent standing wave comprises the ground state at the center of every galaxy, surrounded by an extensive halo of slowly varying density waves that fully modulate the halo density on the de Broglie scale by self-interference \citep{Schive2014, Schive2014b, Mocz2017, Veltmaat2019, Hui2021}. Condensates are inherently non-relativistic, and hence $\psi$DM is viable as dark matter, behaving as CDM on large scales, exceeding the de Broglie wavelength, as demonstrated by the pioneering simulations of \citet{Schive2014}. This $\psi$DM interpretation makes several unique predictions, including a core profile with the unique soliton profile (see section \ref{two}) and a more significant soliton radius, $r_{sol}$, in lower mass galaxies  $m_{gal}$, of lower momentum, as a consequence that the de Broglie wavelength sets the soliton radius.  Following this scaling relation:  $r_{sol}\propto m_{gal}^{-1/3}$ found by \citet{Schive2014b} and verified in independent simulations, \citep{Schwabe2016, Niemeyer2020}.

The effect of tidal stripping in $\psi$DM has been examined recently by \citet{Schive2020, Du2018} demonstrating the resilience of the soliton, which is self-reinforcing so that halo is stripped first, with an abrupt disruption of the soliton predicted to follow if significant tidal stripping of the soliton follows\citep{Du2018}. A steady loss of the halo is found in the $\psi$DM simulation of \citet{Schive2020}, for a typical dwarf of $10^9M_\odot$, following a circular orbit of $\simeq 200$ kpc in a Milky Way sized halo, showing that the halo density drops relative to the soliton over several Gyrs with little change in the slope of the density profile of the halo. This behavior has been recently claimed to bracket the range of profiles found for dwarf spheroidal galaxies, which are observed to follow the predicted form with a clear core and halo structure seen in all the stellar profiles of the classical dwarf spheroidal galaxies \citep{Pozo2020}, and this is also supported by the generally lower level of the velocity dispersion seen in the halo relative to the core as predicted for $\psi$DM \citep{Pozo2020}. This phenomenon is reinforced by \citet{Schive2020} as he already found that tidal stripping increases the density contrast between the core and the halo, as the halo is relatively easily stripped compared to the core. As mass outside the tidal radius is stripped, the core should relax, becoming less massive and hence more extended, obeying the uncertainty principle. With increased stripping such that the tidal radius becomes similar in size to the core radius, the core will be disrupted, perhaps rather abruptly, leaving just extended tails \citep{Du2018}. A recent dedicated $\psi$DM simulation that explores the properties of the Eridanus II dwarf shows that under steady modest stripping, the core remains stable as the halo is reduced in density, leading to an enhanced contrast over time between the relatively dense core and the tenuous surrounding halo\citep{Schive2020},
and the range of core-to-halo variation predicted by this simulation has been shown to match well the family of profiles of dSph galaxies, indicating most are stripped at some level, relative to the more distant ``isolated" dSphs \citep{Pozo2020}.

Here we determine whether this predicted momentum dependence of $\psi$DM can account for the relatively unusual properties of Crater II, given that its relatively wide core and low-velocity dispersion relative to the classical dSphs seems to support this possibility qualitatively. Similarly, it has been proposed that an ultra-light boson of $m_\psi c^2 = ( 0.6 - 1.4 ) \times10^{-22}$ eV accounts for the large size and modest velocity dispersion stars within Antlia II, consistent with boson mass estimates for more massive dwarf galaxies with smaller dark cores \citep{Schive2014} and places Antlia II close to the lower limiting Jeans scale for galaxy formation permitted by the Uncertainty Principle for this very light boson mass. New spectroscopy data has revealed that Antlia II has a systematic velocity gradient that is comparable with its velocity dispersion \citep{Ji2021} indicating that it has suffered tidal elongation or possibly rotating, at least in the outskirts where the sign change of the velocity gradient is apparent\citep{Ji2021}. The tidal field of Crater II is also thought to have had a significant effect given the small pericenter that is now established for Crater II in the latest 'Gaia' based analysis \citep{Ji2021}, though it does not show any clear evidence for a velocity gradient nor visible elongation along its orbit, unlike the case of Antlia II where tidal effects appear more evident \citep{Ji2021}. Moreover, the range of core-to-halo variation predicted by this simulation has been shown to match well the family of profiles of dSph galaxies, indicating most are stripped at some level relative to the more distant ``isolated" dSphs \citep{Pozo2020}. 
 
Significant tidal stripping is now firmly expected for Crater II from its Gaia based orbit, which is confirmed to have a large ellipticity, with a  current radius distance of 117 kpc \citep{Torrealba2016}, well within the Milky Way halo \citep{Sanders2018}, and a pericenter of only $18^{+14}_{-10}$ kpc \citep{Fritz2018,Battaglia2021}, so that that significant tidal stripping is regarded to be likely, perhaps resulting in $ 90\%$ reduction in its mass, with the possible presence of visible tidal distortion\citep{Ji2021}.
 
 The "ultra-light" boson mass solution explored here, with a mass of $\simeq 10^{-22}eV$, is claimed to underestimate the observed amplitude of the Lyman-$\alpha$ power spectrum\cite{Irsic2017} on small scales. This argument relies on an uncertain analogy with previous “Warm" dark matter estimates, rather than employing a self consistent hydrodynamical $\psi$DM simulation that are computationally too demanding currently. We also emphasize empirical evidence that AGN activity may significantly boost the forest power spectrum \cite{Madau2015, Padmanabhan2021}, given the detections of double-peaked Ly$\alpha$ emitters at high redshift, $z\simeq 6$ \citep{Hu2016, Bosman2020, Gronke2021} and the wide ``gaps" in the forest at $z>5$ reported by \citep{Becker2015}, that imply sparsely distributed AGNs\citep{Gangolli2021} at such high redshifts may significantly boost the forest variance in the power spectrum above current standard CDM based predictions. These observations lend support to the proposal that AGN are responsible for the bulk of reionization \citep{Madau2015}, which would imply a different heating history and less uniform reionization, affecting the interpretation of the forest power spectrum, especially on small scales. Moreover, \citet{Zhang2018} points out that the lower mass limit suggested by \citet{Irsic2017} was computed without taking into account the small scale quantum pressure in the $\psi$DM context that sets a Jeans scale inherent to Wave Dark Matter, which limits the formation of small scale structure.

The outline of this paper is as follows; firstly, we describe the radial structure and internal dynamics of the $\psi$DM core in section 2 for comparison with the data. We then describe the tidal effects
in section 3, predicted by recent $\psi$DM simulations, and discuss the possible evolution of Crater II in section 4. Finally, in section 5, we discuss our conclusions regarding the origin of Crater II in the
context of $\psi$DM with significant tidal stripping. 

\section{Comparison of Wave Dark Matter Predictions with observations of Crater II. }\label{two}

Ultralight bosons, such as Axions, provide an increasingly viable interpretation of dark matter as explored in \citet{Widrow1993,Hu2000,Arvanitaki2010,Bozek2015,Schive2014,Hui2017}.
In the simplest case, without any self-interactions, the boson mass is the only free parameter, which if sufficiently light means the de-Broglie wavelength exceeds the mean free path, set by the mean cosmological density of dark matter, and thus satisfies the ground state condition for a Bose-Einstein condensate. In this case the density field is simply described by one coupled Schroedinger-Poisson equation, which in comoving coordinates reads:
\begin{align}
& \biggl[i\frac{\partial}{\partial \tau} + \frac{\nabla^2}{2} - aV\biggr]\psi=0\,,\\
& \nabla^2 V =4\pi(|\psi|^2-1)\,.
\end{align}
Here $\psi$ is the wave function, $V$ is the gravitational potential and $a$ is the cosmological scale factor. The system is normalized to the time scale $d\tau=\chi^{1/2} a^{-2}dt$, and  to the  length scale $ \xi = \chi^{1/4} (m_B/\hbar)^{1/2} {\mathbf x}$, and $\chi=\frac{3}{2}H_0^2 \Omega_0$  where $\Omega_0$ is the current density parameter \cite{Widrow1993}(Check \citet{Schive2014} to understand how the simulations are normalized to the cosmological density where $ \xi$ is used). 

Recently it has proven possible with advanced GPU computing to make the first cosmological simulations that solve the above equations, \citep{Schive2014,Schwabe2016,Mocz2017,May2021}, demonstrating that large scale structure evolves into a pattern of filaments and voids that is indistinguishable from CDM \citep{Schive2014}, as expected for this non-relativitc form of dark matter. However, the virialized halos formed in
the $\psi$DM simulations are very different from CDM, displaying a pervasive self interfering wave structure, with a solitonic core, representing the ground state that naturally explain the dark matter cores of dwarf spheroidal galaxies \citep{Schive2014b}. Surrounding the soliton core, there is an extended halo with a ``granular" texture on the de-Broglie scale, due to interference of excited states, but which when azimuthally averaged follows closely the Navarro-Frank-White (NFW) density profile \citep{Navarro1996,Woo2009,Schive2014,Schive2014b}. This granular interference structure within the halo is predicted to produce noticeable lensing flux anomalies\citep{Chan2020,Hui2021} that are pervasive and hence, statistically unlike the relatively rare sub-halo structure of CDM.

The fitting formula for the density profile of the solitonic core in a $\psi$DM halo is obtained from cosmological simulations \citep{Schive2014,Schive2014b}:
\begin{equation}\label{eq:sol_density}
\rho_c(r) \sim \frac{1.9~a^{-1}(m_\psi/10^{-23}~{\rm eV})^{-2}(r_c/{\rm kpc})^{-4}}{[1+9.1\times10^{-2}(r/r_c)^2]^8}~M_\odot {\rm pc}^{-3},
\end{equation}
where the values of the constants are: $c_1=1.9$, $c_2=10^{-23}$ , $c_3=9.1\times10^{-2}$; $m_\psi$ is the boson mass
and $r_c$ is the solitonic core radius. The latter scales with the product of the galaxy mass and boson mass, obeying the following the scaling relation which has been derived from our simulations \cite{Schive2014b}:
\begin{equation}\label{eq:sol_radius}
r_c=1.6\biggl(\frac{10^{-22}}{m_\psi}  eV \biggr)a^{1/2}
\biggl(\frac{\zeta(z)}{\zeta(0)}\biggr)^{-1/6}
\biggl(\frac{M_H}{10^9M_\odot}\biggr)^{-1/3}kpc,
\end{equation}
Where $a=1/(1+z)$. Beyond the soliton, at radii larger than a transition scale ($r_t$), the simulations also reveal the halo is approximately NFW in form, presumably reflecting the non-relativistic nature of condensates beyond the de Broglie 
scale, and therefore the total density profile can be written as:
\begin{equation}\label{eq:dm_density}
\rho_{DM}(r) =
\begin{cases} 
\rho_c(r)  & \text{if \quad}  r< r_t, \\
\frac{\rho_0}{\frac{r}{r_s}\bigl(1+\frac{r}{r_s}\bigr)^2} & \text{otherwise,}
\end{cases}
\end{equation} 

More explicitly, the scale radius of the solitonic solution, which represents the ground state of the Schrodinger-Poisson equation, is related to size to the halo through the uncertainty principle. From cosmological simulations, the latter is found to hold non-locally, relating a local property with a global one (for more details we refer to \citet{Schive2014b}).

 The classical dwarf galaxies are known to be dominated by DM, and so the stars are treated as tracer particles \cite{Gregory2019,McConnachie2006,McConnachie2006b,Kang2019} moving in the gravitational potential generated by DM halo density distribution.

In this context, the corresponding velocity dispersion profile can be predicted by solving the spherically symmetric Jeans equation:
\begin{equation}\label{eq:sol_Jeans}
\frac{d(\rho_*(r)\sigma_r^2(r))}{dr} = -\rho_*(r)\frac{GM_{DM}(r)}{r^2}-2\beta\frac{\rho_*(r)\sigma_r^2(r)}{r},
\end{equation}
where $M_{\mathrm{DM}}(r)$ is the mass  DM halo obtained by integrating the spherically symmetric density profile in Eq. \eqref{eq:dm_density}, $\beta$ is the anisotropy parameter (see Binney \& Tremaine 2008\cite{Binney2008}, Equation (4.61)), and $\rho_*(r)$ is the stellar density distribution defined by the solitonic wave dark matter profile:
\begin{equation}\label{eq:stellar_density}
\rho_{*}(r) =
\begin{cases} 
\rho_{1*}(r)  & \text{if \quad}  r< r_t, \\
\frac{\rho_{02*}}{\frac{r}{r_{s*}}\bigl(1+\frac{r}{r_{s*}}\bigr)^2} & \text{otherwise,}
\end{cases}
\end{equation}
where,
\begin{equation}\label{eq:stellar_1}
\rho_{1*}(r) = \frac{\rho_{0*}}{[1+9.1\times10^{-2}(r/r_c)^2]^8}~N_* {\rm kpc}^{-3},
\end{equation}

Here, $r_{s*}$ is the 3D scale radius of the stellar halo corresponding to $\rho_{0*}$ the central stellar density, $\rho_{02*}$ is the normalization of $\rho_{0*}$ at the transition radius and the transition radius, $r_t$, is the point where the soliton structure ends, and the halo begins at the juncture of the core and halo profiles. The equivalence of equations \eqref{eq:stellar_density} $\&$ \eqref{eq:stellar_1} with equations \eqref{eq:dm_density} $\&$ \eqref{eq:sol_density} respectively. This behavior we have now established holds generally for the well studied dSph galaxies which all fit well the soliton core profile and with extended halos \citet{Pozo2020}. This behavior may be taken to imply simply that the stars trace the dark matter, as may be expected approximately in dark matter dominated galaxies of spheroidal morphology. Note also that the soliton form is closely similar to the standardly adopted Plummer form widely used to model 
dSph galaxies, where it fits well to several times the core radius but falls well short of the extended stellar halos now commonly found \citep{Torrealba2019,Chiti2021,Collins2021,Pozo2020}.

Finally, the predicted  velocity dispersion profile can be projected along the line of sight to compare with the observations, as presented in Figure~4:
\begin{equation}\label{eq:sol_projected}
\sigma^2_{los} (R) = \frac{2 }{\Sigma_{*}(R)} \int_{R}^{\infty} \biggl(1-\beta \frac{R^2}{r^2}\biggr) \frac{\sigma^2_r(r)\rho_*(r)}{(r^2-R^2)^{1/2}} r dr,
\end{equation}
where,
\begin{equation}
\Sigma_{*}(R) =2\int_{R}^{\infty} \rho_*(r)(r^2-R^2)^{-1/2}rdr\,. 
\end{equation}

\begin{figure*}
	\centering
	\includegraphics[width=0.99\textwidth,height=14cm]{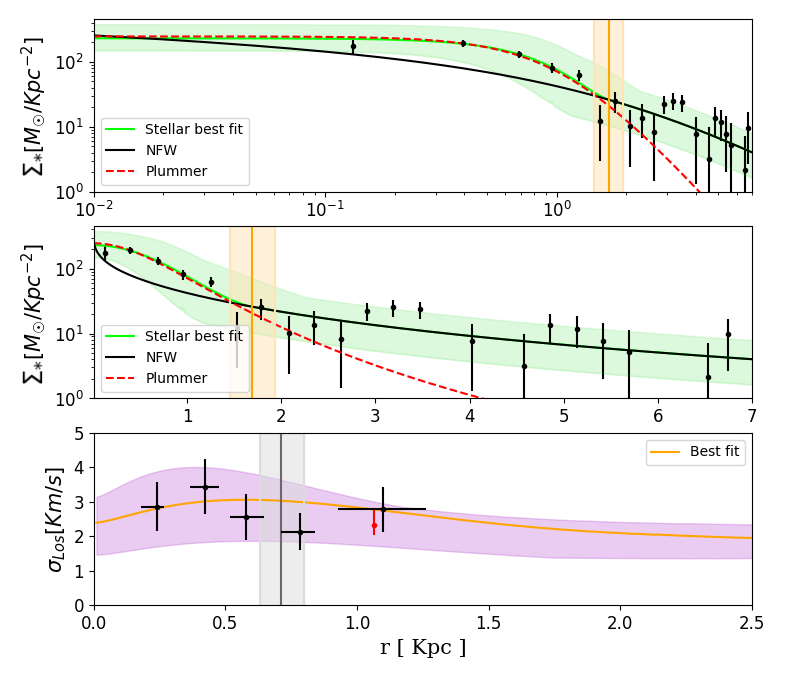}
	\caption{Stellar and velocity dispersion profiles of Crater II compared with $\psi$DM. {\bf Upper panel}: The data points are the star counts of Crater II rebinned from \citet{Torrealba2016}, after applying background subtraction based on their asymptotic limit (red curve in Figure 5 of \citet{Torrealba2016}), compared to the $\psi$DM profile shown as the green shaded area, representing the 2$\sigma$ range of the posterior distribution of profiles, including the soliton core and the outer halo approximate by the NFW form. The apparent excess at $\sim$3kpc seems to be a product of possible contamination by background galaxies\citep{Torrealba2016}. A standard Plummer profile is also shown, which is very similar to the soliton form but underpredicts the observations outside the core. The orange vertical shaded area indicts the transition radius of the $\psi$DM core. {\bf Middle panel}: this shows the same comparison as the upper panel, but on a linear scale, so the extent of the halo can be appreciated better. {\bf Lower panel}: The data points are the velocity dispersion measurements from \citet{Caldwell2017} which are compared to the predicted velocity dispersion for Crater II corresponding to the stellar profile fits to the star counts in the upper and central panels for the 2$\sigma$ range of the MCMC fits (see Fig.\ref{fig6} for details). The vertical grey shaded area indicates the stellar core radius and uncertainty from the above-stellar profile.}\label{fig3}
\end{figure*}

We now apply the above to the newly measured dispersion profile of Crater II dwarf galaxy, discovered by \citet{Torrealba2016} and some ideal $\psi$DM cases. This unusual galaxy was identified in imaging data of the VST ATLAS survey and seemed to be located at $\simeq$120 kpc from the sun. Designed to search for extended low-surface-brightness emission \citep{Torrealba2016}. The galaxy Crater II is one of the largest of the low mass dwarfs orbiting the Milky Way, with a half-light radius $\sim 1.08$ kpc, and also has a relatively low surface brightness dwarfs, similar to cases of Tuc II, Tuc IV, and UMa II. Its stellar velocity dispersion profile has recently been measured with deep spectroscopy by \citet{Caldwell2017} and found to be unusually low with a surprisingly low mean value of only $2.7km/s$ traced to over 1kpc shown in \ref{fig3} where we see the data is consistent with the characteristic $\psi$DM $\sigma$ form which peaks at the soliton radius and declines into the halo (see figure \ref{fig3}).

\begin{table*}
\centering
\begin{tabular}{|c|c|c|c|c|c|c|c|c}
\hline
Galaxy& $r_c$  & $r_{t}$ &  $M(r<r_{h})$ &$M(r<r_{h}),_{obs}$&  $ M_h$& $*_{age,obs}$& $[Fe/H],obs$ &Refs\\
&(kpc)&  (kpc) &$(10^{6}M_\odot)$& $(10^{6}M_\odot)$&  $(10^{8}M_\odot)$& Gyr&&\\
\hline
Crater II  &$0.71^{+0.09}_{-0.08}$ & $1.68^{+0.25}_{-0.23}$  &$7.17^{+12.83}_{-4.56}$&$4.4^{+1.2}_{-0.9}$&$2.93^{+2.99}_{-1.44}$&$\sim10$&$-1.98^{+0.1}_{-0.1}$&Caldwell(2017) \\
\hline
\end{tabular}
\caption{Observations and $\psi$DM profile fits. Column 1: Dwarf galaxy name, Column 2: Core radius  
$r_c$,  Column 3: Transition radius $r_t$, Column 4: Dynamical mass $M(r<r_{h})$, Column 5: Observed dynamical mass $M(r<r_{h}),_{obs}$, Column 6: Halo Mass $M_{h}$, Column 7: Observed Stellar age, Column 8: Observed mean stellar metallicity, Column 9: References of the observed data.}
\label{tabla:1}
\end{table*}

\section{Tidal Effects}\label{three}

Below we describe different effects of the tidal stripping including the mass-loss, core enlargement and the core-halo density transition.

We firstly calculate the relation between $M_H$( halo mass) and $M_c$ (core mass) using Eq. (\ref{eq:sol_density})  until the core radius, as indicated in Eq (\ref{eq:dm_density}). After that, the core mass-loss rate is inferred for each orbit with the following formula \citep{Du2018}:
\begin{equation}\label{eq:dege}
d(M_{c}(t))/dt=-2\times M_{c}(t) e^{a[\frac{3\mu(M_{c}(t))}{2\gamma}]^2+b[\frac{3\mu(M_{c}(t))}{2\gamma}]+c},
\end{equation}
With the  best fitting parameters values: (a,b,c)$\equiv$(5.89794  $10^{-5}$, $-8.72733$  $10^{-2}$, 1.6774), $\gamma$, the effect of the centrifugal force owing to synchronous rotation of the orbiting satellite, assuming it to be a rigid body, fixed to 3/2 for a soliton \citep{Du2018} and $\mu$, the density ratio between the central density of the soliton $\rho_c$ and the average density of the host within the orbital radius $\rho_{host}$, $\mu \equiv \rho_c/\rho_{host} $ \citep{Du2018}. Nevertheless, $\mu$ must be recalculated across the orbits  due to the core density loss along each orbit:

\begin{equation}\label{eq:rc_sol}
\mu=4.38\times10^{10} m_\psi^{6}  M_{c}^{4}d^{3} m_{host}^{-1},
\end{equation}

\begin{figure*}
	\centering
	\includegraphics[width=0.99\textwidth]{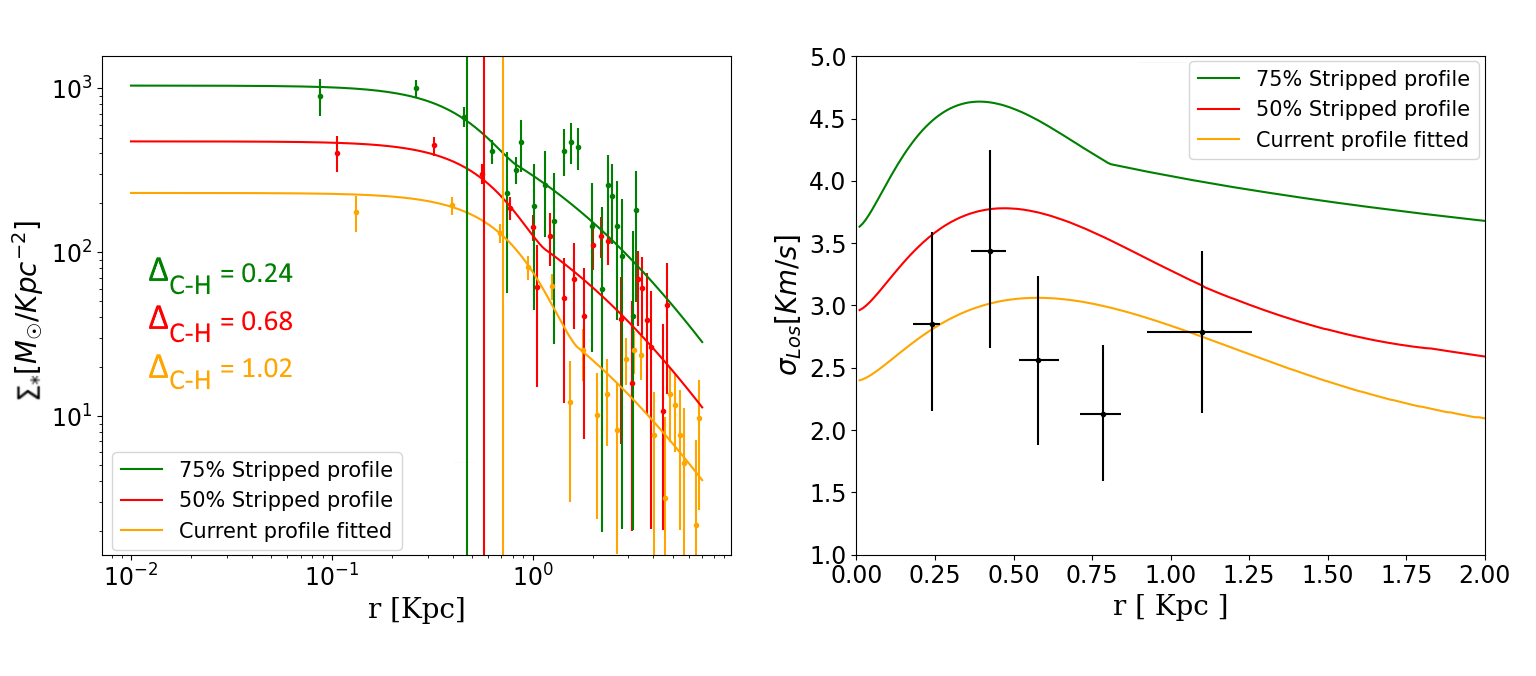}
	\caption{Evolution of the density and velocity dispersion profiles predicted for Crater II. {\bf The left panel}: Possible evolution of Crater II`s stellar profile in a $\psi$DM context, with the measure profile shown in orange, whereas the red and the green represent previous less stripped profiles with $50\%$ and $75\%$ more mass. Notice how the profile becomes "stretched" with a broader core (marked by the vertical lines) and a bigger density gap between the core and the halo (larger $\Delta_{C-H}$).  {\bf The right panel}: Evolution of the velocity dispersion profile due to tidal stripping corresponding to the same epochs as the left panel. The peak of the dispersion moves to larger radius as stripping increases, following the expansion of the core, and the distinction in velocity between the core and the halo diminishes.\label{fig4}}
\end{figure*}

For $\psi$DM, the core and halo are coupled, with the ground state soliton surrounded by the halos of excited states. A Core-halo relationship has been established in the simulations, with more massive solitons formed in more massive halos that are denser because of the higher momentum. Mass loss by tidal forces readily strips the tenuous halo and, in turn, is expected to affect the soliton via the core-halo relation, but in a smaller proportion \citep{Du2018}.

Previously we have found an approximate proportionality between $ r_c $ and the observed half light radius, $ r_h $, for local dSph galaxies in \citet{Pozo2020}, of about $ \simeq $ 1.3. This $ r_h/r_c $  ratio was has also been pointed out by \citet{Lazar2020, Schive2014, Schive2014b}. So as well as computing the mass loss due to tidal stripping using Eq (\ref{eq:dege}), we continuously increment $r_c$ by updating the mass-loss in  Eq. (\ref{eq:sol_density}). Finally, we use the above ratio between $r_c$ and $r_h$ to increment $r_h$ across the orbits (see righthand panel of Fig.\ref{fig5}). The enlargement of $r_t$,  (extracted from 
the simulations made by \citet{Schive2020}, where it is defined in which ratio should the transition radius of a Milky way's  satellite dwarf galaxy increase due to Milky Way's tidal forces: "The halo surrounding the central soliton is
found to be vulnerable to tidal disruption; the density
at r > $r_t$ decreases by more than an order of magnitude after $\sim$ 2 Gyr") will explain the observed $\Delta_{C-H}$( density drop between the core and the halo, $\Delta_{C-H}= \log{\rho_{C}/\rho_{H}}$, where $\rho_{C}$ is the asymptotic central core stellar density and $\rho_{H}$ is the stellar density at the transition radius($r_t$) ) changes in the halo of the galaxies.

\section{Model Results}\label{four}

\begin{figure*}
	\centering
	\includegraphics[width=0.99\textwidth]{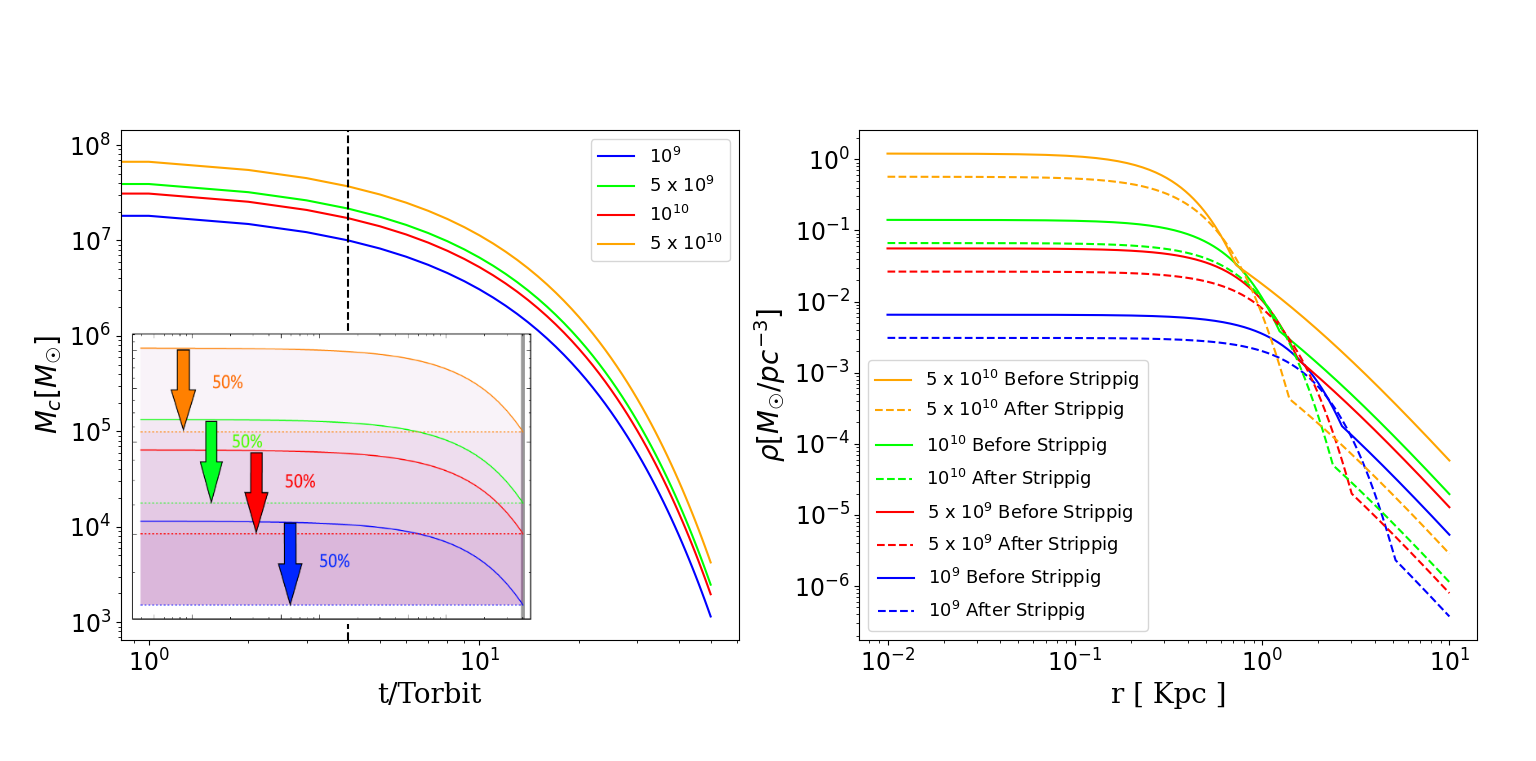}
    \caption{The evolotion of Four typical dSph $\psi$DM density profiles predicted for orbits within the Milky Way, with a constant density ratio of; $\mu$=50. {\bf The left panel}: Core mass loss of the four profiles. The vertical dashed line marks four orbits. The small panel shows in detail the profiles after each has lost $50\%$ of the total mass to steady tidal stripping after the first four orbits. {\bf The right panel}:Tidal evolution of the $\psi$DM density profiles. The thick lines represent the original profiles before the stripping process, while the dashed lines represent their situation after four orbits. Notice how in all the cases, the core becomes extended due to losing mass as the transition radius increases.}\label{fig1}
\end{figure*}

\begin{figure}
	\centering
	\includegraphics[width=0.5\textwidth,height=7cm]{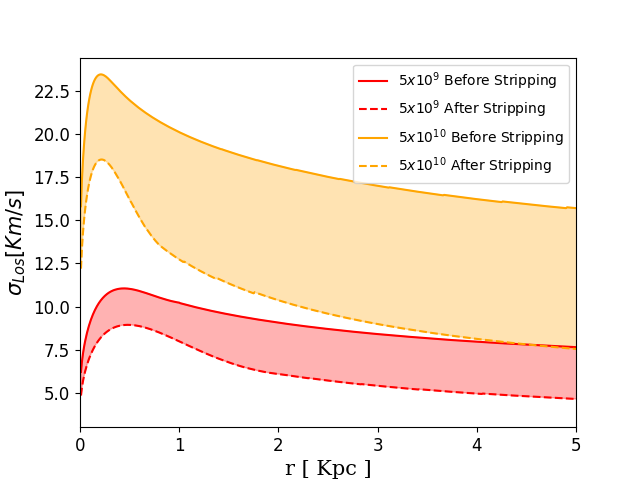}
	\caption{Velocity dispersion profile evolution for two tidally evolved profiles of different galaxy mass spanning the classical dwarf range, as in figure \ref{fig1}. Notice how the core is more evident in the dispersion profile for the more concentrated, massive soliton, indicated by the orange profile, compared to the softer core-halo transition for the less massive dwarf indicated by the red profile. }\label{fig2}
\end{figure}

\begin{figure*}
	\centering
	\includegraphics[width=0.99\textwidth,height=8cm]{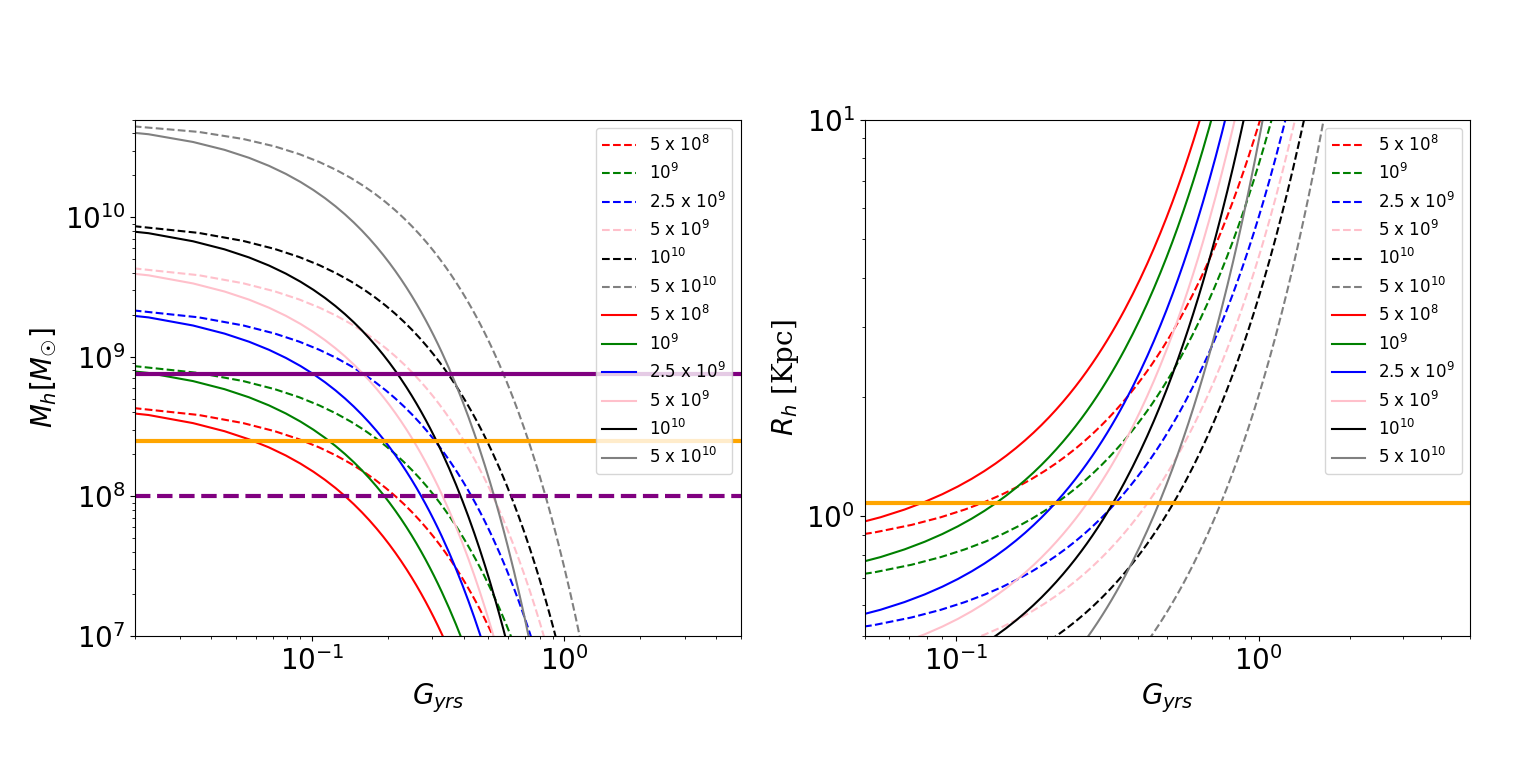}
	\caption{Evolution of mass and $r_{h}$ radius due to tidal stripping. The thick  solid and dashed lines are with a Milky way host of $5\times10^{12}$ $M_\odot$ and  $10^{12}$ $M_\odot$, respectively. {\bf The left panel}:nras
	Total mass loss evolution predicted for Crater II in the $\psi$DM context. The orange horizontal line represent the best fit mass of Crater II (see figure \ref{fig3} and table \ref{tabla:1}). The solid and dashed purple horizontal lines indicates the maximum and minimum allowed masses of Crater II from our analysis, corresponding limiting purple contours of Figure 1. {\bf The right panel}: Predicted evolution of the $r_{h}$  radius growth for Crater II, in a $\psi$DM context. The orange horizontal line represent Crater II's actual $r_{h}$.}\label{fig5}
\end{figure*}

\begin{figure*}
	\centering
	\includegraphics[width=0.99\textwidth,height=11cm]{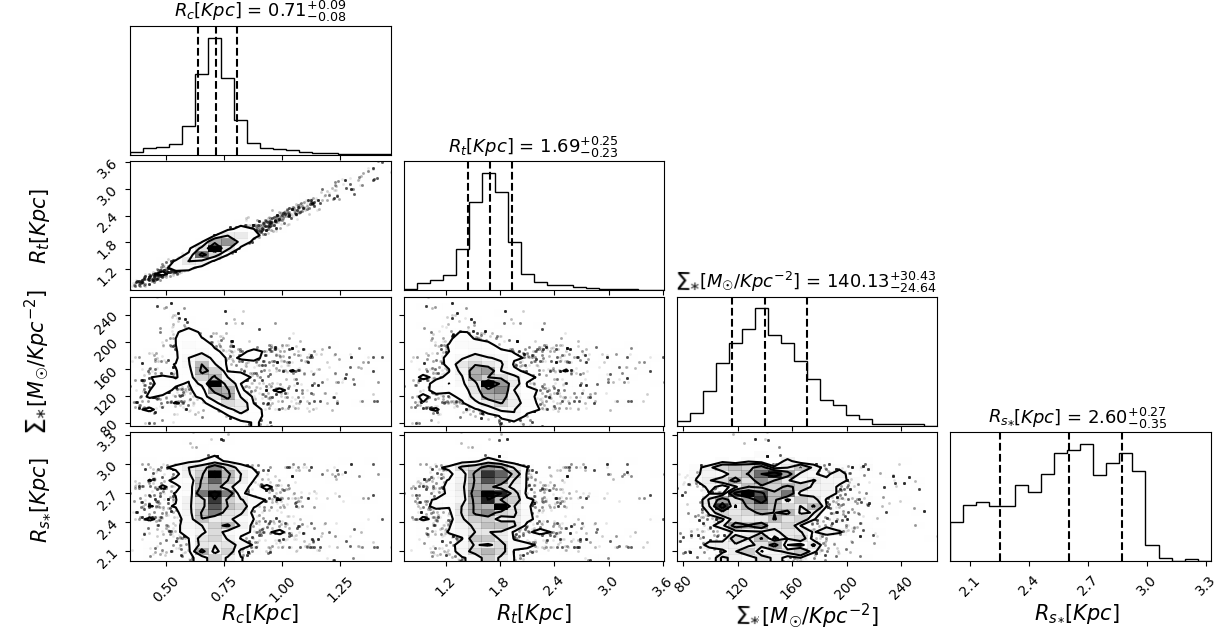}
	\caption{Crater II: correlated distributions of the free parameters.  As can be seen the core radius and transition radius are well defined despite the flat input priors, indicating a reliable result. The contours represent the 68\%, 95\%, and 99\% confidence levels. The best-fit parameter values are the medians(with errors), represented by the dashed black ones, and tabulated in Table\ref{tabla:1}.}\label{fig6}
\end{figure*}

\subsection{Crater II}\label{fourtwo}

Here we compare the measured velocity dispersion and the stellar profile of Crater II with $\psi$DM. We fit the data with the following free parameters; the core radius $r_c$ of the soliton profile given by Eq.\eqref{eq:stellar_1}, the transition radius $r_{t}$ between the core and the halo, the central 3D stellar density $\rho_{0*}$ and the 3D scale radius of the stellar halo $r_{s*}$  describing the scale radius of the NFW-like halo that we infer from fitting to the outer stellar profile beyond the transition radius. Note that the boson mass is fixed, with a value of $1.5\times10^{-22}$ eV, consistent with our previous dynamical work on dwarf galaxies in the context of $\psi$DM \citet{Schive2014,Schive2014b,Broadhurst2020,Pozo2020}.

 In generating our model profiles, we solve the spherically symmetric Jeans equation, described above, Eq. \eqref{eq:sol_Jeans}, subject to a total mass for Crater II of $2.93^{+2.99}_{-1.44}\times 10^{8}M_\odot$, that we obtain from  fitting the core radius size ( Eq.\eqref{eq:stellar_1} ) in the stellar profile data with a fixed boson mass of $1.5\times10^{-22}$ eV, which is consistent with the dynamical mass estimated by \citet{Caldwell2017} (see table (\ref{tabla:1}) .The transition radius, $r_t$, is expected to be two or three times larger than the core radius in simulations of $\psi$DM \cite{Schive2014,Schive2014b}, which we show below is consistent with $r_t\sim$2.4$r_c$ that we derive here for Crater II.


The results are listed in table (\ref{tabla:1}), and figure (\ref{fig3}) shows the self-consistency of these data in both stellar and kinematic profiles. The transition radius, $r_{t}$ at which the profile changes from being dominated by the halo rather than the soliton is marked with a vertical orange line in Figure (\ref{fig3}). The green and purple shadowed areas represent the 2$\sigma$ range, respectively. The comparison shown in Figure (\ref{fig3}) between the models and the data represents a consistency check, where we simply employ the Jeans equations in making predictions for the velocity dispersion profile (purple model band in lower panel of Figure (\ref{fig3})  when inputting the set of core+halo profiles that acceptably fit the stellar profile data (green model band Figure (\ref{fig3}) and subject to a constraint on the total galaxy mass set by the mean level of velocity dispersion of about 3km/s. Note, the limited precision of the velocity dispersion data does not yet  warrant a classic combined Jeans analysis.

Figure (\ref{fig4}) shows our predicted evolution of the stellar profile of Crater II as well as the change in the velocity dispersion profile for two choices of galaxy mass and spanning a mass loss of up to 50\%. As a consequence of continued core mass loss \citep{Du2018}, a widening of the core is induced, described by Eq. \eqref{eq:sol_radius}. It is important to point out how both densities of core and halo seem to decrease, with the halo changes more strongly over time, in good agreement with the halo's greater weakness against tidal forces \citep{Schive2020}. This tidally induced mass loss results in a reduction of the velocity dispersion while widening the core so that the dispersion profile has a less pronounced peak that shifts the larger radius, reflecting the widening core as tidal stripping proceeds as seen in Fig.\ref{fig5} for the parameters of Crater II. This is in line with the simulations of \citet{Fu2019} where prolonged tidal stripping should produce a drop in the mean velocity dispersion in conjunction with a  half-light radius increasing \citep{Fattahi2018,Sanders2018,Torrealba2019}. 

We conclude that the extended core size of $r_c \simeq 0.71$ kpc of Crater II is is the product of its low halo mass($\simeq 10^8$) and significant tidal stripping of the halo have have occurred which is natural in the $\psi$DM context, while CDM struggles to explain the observed combination of low-velocity dispersion and large radius\citep{Fattahi2018}. Moreover,  Fig.\ref{fig3}  clearly show how a cuspy NFW profile is unable to explain the stellar density and kinematic behavior in the core, under the assumption that stars trace the dark matter. We also note that we have not adopted the commonly used Plummer profile for the stellar profile, preferring instead the soliton form that fits well the stellar profile of Crater II (Eq. \eqref{eq:stellar_density}) derived in section \ref{two}.

\subsection{General case}\label{fourone}
Here we predict the tidal evolution of dwarf galaxy profiles in the context of $\psi$DM. Figure (\ref{fig1}) shows how the soliton core mass should decrease with time over several Gyrs, according to Eq.(\ref{eq:dege}). The soliton profile remains unchnaged until it becomes stripped, then the core relaxes into a softer profile during this process, (see figures (\ref{fig1}) and (\ref{fig2}). This right hand panel of Figure (\ref{fig1}) shows the soliton becoming wider, indicated by the difference between thick and dashed lines of the same color, due to the halo mass loss described by Eq(\ref{eq:sol_radius}) and at the same time the amplitude of the density gap and the transition radius can be seen to increase as the halo is stripped. Similar behaviour is noticeable in the recent $\psi$DM simulations of \citep{Schive2020}, where tidal stripping has been approximated for the orbiting dwarf Eridanus II.

Figure (\ref{fig2}) shows illustrative velocity dispersion profiles for a range of $\psi$DM mass profiles highlighting the transition from the soliton core to the outer NFW-like outer profile \citep{Schive2014,Schive2014b,Vicens2018}. The velocity dispersion profiles are listed in the right upper panel and cover one order of magnitude in the total mass starting from $5 x 10^{9} M_{\odot}$ to $5 x 10^{10} M_{\odot}$. Solid and dashed lines differentiate the original ideal isolated profile from the stripped one for each system. In terms of the Jeans based calculation of the dispersion profile, the choice of $\beta$ is not very important, affecting the velocity dispersion well within the core, where is rises (or fall) sharply for a positive (or negative) value of beta and remains flat if isothermal. The data appear to favor a mildly negative value for $\beta$ as the dispersion of the innermost bin is lower than the mean, though quite uncertain, as can be seen in Figure 1 (lower panel), consistent with our adopted $\beta=-0.5$ and this is similar to the value chosen by \cite{Caldwell2017} for modeling dwarf spheroidals, to counter the rising dispersion profile that would otherwise result from the “cusp” of an NFW profile. The main influence on the velocity dispersion profile is from the presence of the soliton core, rather than the choice of $\beta$, because the soliton is dense relative to the halo (by a factor of about 30), generating a peaked form  at about the core radius, as can be seen in Figure 4 and also in figure 1.

\section{Discussion and Conclusions}\label{five}

 The existence of Crater II, with its unusually large size and relatively low-velocity dispersion, has been a surprise as it strains the very definition of a "dwarf" galaxy with its large size and also because is it in significantly greater tension with  $\Lambda$CDM
 than the dSph galaxies \citep{Amorisco2019,Borukhovetskaya2021}, for which sizeable dark cores are often claimed, at odds with the 
 standard prediction for collisionless CDM particles (or black holes) that low mass galaxies should be small and dense rather than large and diffuse like Crater II.
 
 Instead, in this paper, we have shown that Crater II can be readily understood in the context of dark matter as a Bose-Einstein condensate by comparison with the profiles of galaxies generated in $\psi$DM simulations \citep {Schive2020}, especially if tidal stripping is included. One of the most distinguishing features of $\psi$DM is the prominent soliton core. This is quite unlike the usual smooth cores generally explored in other models, where the density
 turns over towards the center becoming constant, but instead, the solitonic core of $\psi$DM is {\it prominent}, raised in density above the surrounding halo by a factor of $\simeq 30$, representing the ground state. Furthermore, the soliton density profile is characterized only by a radius set by the de Broglie Wavelength, with the same shape, irrespective of boson mass, scaling only with the momentum that sets the de Broglie wavelength, so that the soliton is smaller in radius and more massive for higher mass and more concentrated galaxies.
 
 It should be emphasized that the simulations also make clear that there is a marked transition between the core and the halo, and we can see that Crater II does possess a well-defined core with such a visible transition in its stellar profile at a radius of $\simeq 0.7$kpc, shown in Figure \ref{fig3}, and this is despite the relatively low surface brightness of  Crater II, for which the star counts are much lower than typical well studied dSph galaxies. The existence of this core is also supported by the velocity dispersion profile of Crater II, which we have shown is consistent with being peaked at about the stellar core radius, as we have shown is predicted for $\psi$DM in Figures \ref{fig1} \& \ref{fig2}.
 
 Deeper imaging and spectroscopy for Crater II can help considerably in clarifying the level of 
 correspondence between the stellar profile and internal dynamics, in particular, it will be very helpful to see whether the velocity dispersion falls in the halo region beyond the current limit of $r < 1.2$ kpc covered by existing dynamical measurements, where we predict the dispersion to be significantly lower than in the core for the $\psi$DM profile, and quite unlike the rising profiles predicted by CDM for Crater II, which must be assumed to be hosted by a relatively massive galaxy of low concentration in the context of CDM \citep{Amorisco2019,Borukhovetskaya2021}. 
 
 We have also pointed out that the observed stellar profile behavior of Crater II is continuous with the distinctive core-halo structure that appears to be a general feature of the classical dSph galaxies, established in our earlier work \citet{Pozo2020}, where we found that essentially all the well studied dSph galaxies have a prominent stellar core that accurately matches the unique soliton form, and also that the velocity dispersion profiles of these dSph galaxies generally peak near the stellar core radius and are lower in the halo. However, despite this qualitative similarity between Crater II and the dSph class, there is a clear difference in that the dSphs are about a factor 
 three smaller, with a mean core radius of $\simeq 0.25$Kpc, compared to $\simeq 0.7$kpc for Crater II, and also in terms of the characteristic velocity dispersion which is about three times greater for the dSphs, with a mean level of $8-12$km/s compared to only $2.7$km/s for Crater II significantly higher than Crater II, and also of course, as the name "feeble giant" suggests, Crater II is relatively more extended and of unusually low surface brightness than typical dSph galaxies \citep{Torrealba2016}. 
 
 Our analysis of Crater II has examined the possibility that tidal stripping in the context of WaveDM may account for the rather extreme properties of this dwarf galaxy, with its relatively large size and low velocity dispersion. In support of this  we show the stellar profile of the Crater II (upper panel Figure 1) is well fitted by a soliton core plus a shallow NFW halo and that this combined profile, that is generic to WaveDM, is consistent with the extended, shallow, velocity dispersion profile of Crater II, (lower panel Figure 1) where the low mean level of velocity dispersion of $\simeq 3$km/s corresponds to a total galaxy mass of about $3\times10^{8}M_\odot$. The consistency we find here may indicate that the stellar core and DM core are similar in scale, though this is by no means conclusive at the current limited precision of the dispersion profile and relies to some extent the level of velocity anisotropy assumed, but may imply the stars behave essentially as tracer particles within the dark matter dominated potential. Detailed hydrodynamical simulations of gas and star formation within dwarf galaxy haloes will be required for a more definitive exploration of this relationship between stars and DM, that must include the relaxation effects understood to be significant in randomly deflecting stars orbiting through the de Broglie scale density fluctuations predicted for WaveDM halos \citep{Schive2014} and proposed as possible explanation for the increasing scale height of disk stars in the Milky Way with stellar age, by \cite{Church2019}, and \cite{BarOr2019}. For now we content ourselves with the largely qualitative conclusion that tidal stripping of a DM dominated dwarf galaxy in the context of Wave DM can plausibly result in the unusual properties of Crater II,  with an expansion of its soliton core in response to stripping of its DM halo, an effect that follows fundamentally from the Uncertainty Principle for WaveDM.
 
 We have been able to understand these differences in the context of $\psi$DM, as the possible consequence of tidal stripping.  It is
now understood that Crater II has a small pericenter within the Milky
Way, so that tidal stripping should be significant and this we have
shown provides an interpretation of Crater II as a stripped dSph galaxy. This possibility follows
 directly from considering how the soliton expands as the halo mass is stripped away, as proposed by \citet{Du2018} that is implied by the existence of a relatively clear relationship established in the $\psi$DM simulations between the mass of a galaxy and the 
 soliton core, which may act together with the strict inverse relationship required for a soliton (by the Uncertainty Principle), such that as galaxy mass is reduced by tidal stripping, the momentum associated with the soliton ground state is also lower and hence the soliton expands as the de Broglie Wavelength is larger following the inverse
 soliton mass-radius relation, with a reduce soliton density and hence lower velocity dispersion. We may conclude that in order for Crater II to have originated as a typical dSph, its core has expanded by approximately a factor of 2-3, and hence the core mass was 2-3 times higher and hence initially, the total mass would have been $\simeq 10-30\times$ 
 larger, assuming the core-halo mass scaling relation scaling is followed, i.e. $m_{sol}$ $\propto$ $m_{gal}^{1/3}$. We can come to the 
 same quantitative conclusion by comparing the velocity dispersion $\simeq 3$km/s of Crater II, which also differs by a factor of 3 with the typical $8-10$km/s peak dispersion for the 
 dSph's. This agreement is quite compelling 
 for the $\psi$DM interpretation as this factor difference would not be expected, whereas for $\psi$DM it is a requirement as the uncertainty principle dictates that the soliton obeys $r_{sol}\sigma_{sol}$=$h/{8\pi m_\psi}$, (taking $2r_{sol}$ as the width of the soliton wave packet) providing an approximate boson mass of $2.51^{+0.68}_{-0.51}\times10^{-22}$eV, and because 
 $\sigma_{sol}$ and $r_{sol}$ vary inversely with a fixed product, this boson mass estimate is expected to be independent of tidal evolution, in the absence of extreme tidal disruption that may deform the soliton \citep{Du2018}. This value of $m_\psi$ is consistent with estimates for the cores of classical dSph galaxies, $\simeq 1-2 \times 10^{-22}$eV\citep{Chen2017}, and thus Crater II quantitatively reinforces the light boson solution for dark matter. 
 
 
 \section*{Acknowledgements}

TJB thanks Gabriel Torrealba, Tzihong Chiueh and Simon White for useful conversations. Razieh Emami acknowledges the support by the Institute for Theory and Computation at the Center for Astrophysics. George Smoot is grateful to the IAS at HKUST for generous support. Also AP is thankful for the continued support of the DIPC graduate student program. This work has been supported by the Spanish project PID2020-114035GB-100  (MINECO/AEI/FEDER, UE). We thank the referee for a thorough reading and useful suggestions that have improved the paper.

\section*{Data Availability}

The data underlying this article will be shared on reasonable
request to the corresponding author.

\bibliographystyle{mnras}
\bsp	
\label{lastpage}
\end{document}